\documentclass[prl,superscriptaddress,twocolumn, showpacs,preprintnumbers]{revtex4}

\usepackage{natbib}

\usepackage{amssymb}
\usepackage{amsthm}
\usepackage{amsmath}
\usepackage{amsfonts}

\usepackage[mathscr]{eucal}
\usepackage{bm}             

\usepackage{graphicx}
\usepackage{psfrag}
\usepackage{subfigure}

\graphicspath{{.}{.}}

\newcommand{\nextparagraph}{\parskip2ex}

\newcommand{\Phiret}{\Phi_{\mathrm{ret}}}
\newcommand{\Phiadv}{\Phi_{\mathrm{adv}}}

\newcommand{\Kret}{K_{\mathrm{ret}}}
\newcommand{\Kadv}{K_{\mathrm{adv}}}

\newcommand{\barr}{\bar{r}}
\newcommand{\barv}{\bar{v}}
\newcommand{\barmu}{\bar{\mu}}

\theoremstyle{plain}

\theoremstyle{remark}

\begin{document}

\title{Helically symmetric \textit{N}-particle solutions in scalar gravity}

\author{Robert Beig}
\email{Robert.Beig@univie.ac.at}
\affiliation{Gravitational Physics, Faculty of Physics, University of Vienna,
  A-1090 Vienna, Austria}

\author{J.\ Mark Heinzle}
\email{Mark.Heinzle@univie.ac.at}
\affiliation{Gravitational Physics, Faculty of Physics, University of Vienna,
  A-1090 Vienna, Austria}

\author{Bernd G.\ Schmidt}
\email{Bernd.Schmidt@aei.mpg.de}
\affiliation{Max-Planck-Institute for Gravitational Physics, Albert-Einstein-Institute,
  D-14476 Golm, Germany}

\begin{abstract}


Within a scalar model theory of gravity, where
the interaction between particles is given
by the half-retarded $+$ half-advanced solution
of the scalar wave equation,
we consider an $N$-body problem:
we investigate configurations of $N$ particles which
form an equilateral $N$-angle and are in helical
motion about their common center.
We prove that there exists a unique equilibrium
configuration and compute the equilibrium radius
explicitly in a post-Newtonian expansion.

\end{abstract}
\maketitle


Self-gravitating systems with helical symmetry have recently
attracted considerable interest in the context of the numerical
evolution of coalescing neutron star and black hole
binaries~\cite{Yoshida/etal:2006,Klein:2004}. Since numerical codes
simulating relativistic collapse cannot evolve for large times,
initial data should be close to the final plunge. A reasonable
approximation to such initial data might be to consider data given
by spacetimes with a helical Killing vector, the reason being that
gravitational radiation tends to reduce the eccentricity of orbits.
However, not only for numerical purposes, but also from a more
systematic point of view, spacetimes with a helical Killing vector
form a class of time independent solutions of the Einstein equations
which are interesting in themselves and about which little
--- including their existence --- is known. The simplest class of
examples are $N$ point particles of equal mass $m$ in Newtonian
theory forming an equilateral $N$-angle and uniformly rotating about
their common center. We construct the analog of these solutions in a
special relativistic scalar theory of gravity, with particles
interacting via the half-retarded $+$ half-advanced ("symmetric")
solution of the wave equation. The case $N = 2$ (and allowing for
different masses) has been considered by \cite{Bruneton:2006,
Friedman/Uryu:2006}; the electromagnetic $N=2$ case has been treated
in the seminal paper~\cite{Schild:1963}. In this paper we perform a
careful study of the symmetric interaction of particles in helical
motion which is absent in the literature even in the antipodal
($N=2$) case. This serves the purpose of proving the following
result: given $m$ and the angular velocity $\Omega$ of the helical
motion, there exists, like in Newtonian theory, exactly one radius
$\barr_{\mathrm{e}}$, for which the symmetric interaction is
balanced by the centrifugal force.

Although scalar theories of gravity are known to disagree with
experiment, they provide simple test models for relativistic
gravity, mainly because they have just one degree of freedom,
see~\cite{Shapiro/Teukolsky:1993}. Scalar theories derive from an
action
\[
S = \frac{1}{2} \int_M g^{\alpha\beta} \Phi_{,\alpha} \Phi_{,\beta} d^4 x +
4 \pi \int_M \rho \,F(\Phi) d^4 x,
\]
where $(M, g_{\alpha\beta})$ is Minkowski space and $\rho$ the
energy density of matter; we have set $G=1$, $c=1$. The resulting
equation for $\Phi$ is $\Box \Phi = 4 \pi \rho F^\prime(\Phi)$. When
$F(0)$ is chosen to be one, the associated theory has the correct
Newtonian limit. The choice $F(\Phi)= \mathrm{exp} \,\Phi$
corresponds to the model theory recently proposed
in~\cite{Shapiro/Teukolsky:1993}. To further simplify matters, we
make the choice $F(\Phi) = 1 + \Phi$, which corresponds to a
first-order expansion, see~\cite{Giulini:2006,Ravndal:2004}; this
leads to the linear wave equation for $\Phi$,
\begin{equation}\label{waveeq}
\Box \Phi = 4 \pi \rho\:.
\end{equation}

We consider a family of $N$ structureless point particles
of equal mass $m$; let $\bar{x}_n(s_n)$ be the world line
of the $n$\raisebox{0.6ex}{{\footnotesize th}} particle, where
$s_n$ denotes proper time; then
\[
\rho (x) = m \sum_{n = 0}^{N-1} \int
\delta^{(4)}\big(x - \bar{x}_n(s_n)\big)\:d s_n\:,
\]
so that the particle equations of motion are
\begin{equation}\label{equicond}
m \,\frac{d}{d s_n}\!\left[\big(1 + \Phi|_{\bar{x}_n}\big)\,\dot{\bar{x}}_{n}^{\,\alpha} \right]+
(\partial^\alpha \Phi)\big|_{\bar{x}_n} = 0
\end{equation}
for $n = 0,1,\dots (N-1)$,
where the field $\Phi|_{\bar{x}_n}$ acting on the $n$\raisebox{0.6ex}{{\footnotesize th}} particle
is the symmetric solution of~\eqref{waveeq} generated by the remaining particles~%
\footnote{A rigorous justification for this notion of an $N$-particle solution
other than the standard derivation \textit{\`a la} Lorentz-Dirac requires
the point particle limit to be taken within a consistent treatment of fluids
or elastic bodies [Beig, Schmidt, in preparation].}.

In helical symmetry, fields 
are invariant under the action of
a helical Killing vector $\xi$, whose components are $\xi_t = 1$, $\xi_\phi = \Omega = \mathrm{const}$,
$\xi_r = 0 = \xi_z$, and whose Lorentz norm is $\xi^2 = -1 + \Omega^2 r^2$; here, $(r,\phi,z)$
are cylindrical coordinates associated with $\vec{x}= (x,y,z)$.
Helical motion is motion tangent to the Killing orbits,
i.e., circular motion with constant
angular velocity $\Omega$ in planes $z=\mathrm{const}$.
When we define $\mu = \phi - \Omega t$, we find
that a field $\psi$ on $(M,g_{\alpha\beta})$ is helically symmetric, if
it is of the form $\psi(\mu,r,z)$, where $\psi$ is periodic in $\mu$ with
period $2\pi$; helical motion is any motion with $(\mu,r,z)(s) \equiv \mathrm{const}$.

\nextparagraph
\paragraph{Helical solutions of the wave equation.}

We consider the wave equation~\eqref{waveeq}
for a helically symmetric source
$\rho(t, \vec{x}) = \rho_{\mathrm{h}}(\mu,r,z)$,
where $\rho_{\mathrm{h}}$
is $2\pi$-periodic in $\mu$;
we assume $\rho_{\mathrm{h}} = 0$ for $r\geq \Omega^{-1}$
so that the source is confined within the light cylinder $r = \Omega^{-1}$,
where velocities are less than the speed of light.
The retarded solution $\Phiret(t,\vec{x})$ and the
advanced solution $\Phiadv(t,\vec{x})$ of~\eqref{waveeq}
are given by
\begin{equation}\label{Phiretadv}
\Phi_{\mathrm{ret}/\mathrm{adv}} = -
\int_{\mathbb{R}^3}
\frac{\rho(t \mp | \vec{x} -\vec{x}^\prime |, \vec{x}^\prime)}{ | \vec{x} -\vec{x}^\prime |}
\: d^3 x^\prime \:.
\end{equation}
The solutions 
$\Phiret$ and $\Phiadv$
share the symmetries of the source, i.e.,
$\Phiret = \Phiret(\mu,r,z)$. 
To make this explicit, we first
introduce cylindrical coordinates $(t,r,\phi,z)$ and $(r^\prime, \phi^\prime, z^\prime)$ associated
with $(t,\vec{x})$ and $\vec{x}^\prime$ in~\eqref{Phiretadv}; we then find that
the integrand
can be regarded as a $2\pi$-periodic function of $\sigma = \phi- \phi^\prime$,
which further entails that
\[
\Phiret = -
\int\limits_0^{2\pi}\! d\sigma \!
\int\limits_0^{\Omega^{-1}}\!r^\prime d r^\prime \!\!\!
\int\limits_{-\infty}^{\infty} \!\! d z^\prime\:
\frac{\rho_{\mathrm{h}}(\mu - \sigma + \Omega | \vec{x} -\vec{x}^\prime |, r^\prime, z^\prime)}{ | \vec{x} -\vec{x}^\prime |}\:,
\]
where $| \vec{x} -\vec{x}^\prime |^2 = r^2 + r^{\prime\,2} -2 r r^\prime \cos \sigma + (z - z^\prime)^2$.
Consequently, $\Phiret = \Phiret(\mu,r,z)$ with $2 \pi$-periodicity in $\mu$.
Note that the advanced solution $\Phiadv(\mu,r,z)$
arises from $\Phiret(\mu,r,z)$ by 
making the replacement $\Omega \rightarrow (-\Omega)$.

We proceed by
defining a variable $\mu^\prime$ via
\begin{equation}\label{muprime}
\mu^\prime  = \mu - \sigma +
\Omega \left[r^2 + r^{\prime\,2} -2 r r^\prime \cos \sigma + (z - z^\prime)^2\right]^{\frac{1}{2}} \!.
\end{equation}
For fixed $\mu$, $r$, $r^\prime < \Omega^{-1}$, $z$, $z^\prime$, the map
$\sigma \mapsto \mu^\prime$ is monotonically decreasing and thus a diffeomorphism.
This follows from a straightforward computation, where we
invoke de l'Hospital's rule (for $r = r^\prime$, $z = z^\prime$, $\sigma = 2 k \pi$, $k\in\mathbb{Z}$).
Performing a change of integration variables from $\sigma$ to $\mu^\prime$,
where we use that a
shift by $2 \pi$ in $\sigma$ causes a shift by $-2\pi$ in $\mu^\prime$,
we eventually arrive at
\[
\Phiret =  \Omega
\int\limits_0^{2\pi}\! d\mu^\prime \!
\int\limits_0^{\Omega^{-1}}\!r^\prime d r^\prime \!\!\!
\int\limits_{-\infty}^{\infty} \!\! d z^\prime\:
\frac{\rho_{\mathrm{h}}(\mu^\prime,r^\prime,z^\prime)}{\mu - \mu^\prime - \sigma + \Omega^2 r r^\prime \sin\sigma}\,.
\]
In this integral, $\sigma$ is to be regarded as a function
of the other variables, implicitly given by~\eqref{muprime}.
In fact, $\sigma = \sigma(\mu-\mu^\prime,r,r^\prime,z-z^\prime)$,
where $\sigma(\mu,r,r^\prime,z)$ satisfies
\begin{equation}\label{retangle}
\mu - \sigma + \Omega \left[ (r - r^\prime)^2 + 4 r r^{\prime} \sin^2 \frac{\sigma}{2} + z^2 \right]^{\frac{1}{2}} = 0\:.
\end{equation}
We call $\sigma(\mu,r,r^\prime,z)$ 
the retarded angle
associated with $\mu$ (and the particular choice of $r$, $r^\prime$, $z$).
The following properties of $\sigma$ are immediate from the above discussion:
\begin{equation}\label{sigmaperiod}
\frac{d\sigma}{d\mu} > 0 \:, \quad\: \sigma(\mu + 2\pi,r,r^\prime, z) = \sigma(\mu,r,r^\prime,z) + 2 \pi\:.
\end{equation}

Finally, regarding the retarded solution $\Phiret(\mu,r,z)$
as the convolution of an integration kernel
with the source $\rho_{\mathrm{h}}(\mu^\prime,r^\prime,z^\prime)$
yields the so-called retarded kernel
\begin{equation}\label{Kret}
\Kret(\mu,r,r^\prime,z) =
\Omega \;\frac{1}{\mu - \sigma + \Omega^2 r r^\prime \sin \sigma}\:,
\end{equation}
where $\sigma = \sigma(\mu,r,r^\prime,z)$.
As a consequence of~\eqref{sigmaperiod},
$\Kret(\mu,r,r^\prime,z)$ is $2 \pi$-periodic in $\mu$.

As noted above, the advanced kernel $\Kadv(\mu,r,r^\prime,z)$ is given in analogy to~\eqref{Kret},
where $\Omega \rightarrow (-\Omega)$ and $\sigma \rightarrow \sigma_{\mathrm{adv}}$;
\[
\mu - \sigma_{\mathrm{adv}} -
\Omega \left[ (r - r^\prime)^2 + 4 r r^{\prime} \sin^2 \frac{\sigma_{\mathrm{adv}}}{2} + z^2 \right]^{\frac{1}{2}} = 0\:.
\]
From~\eqref{retangle} we conclude that $[-\sigma(-\mu,r,r^\prime,z)]$ satisfies this equation,
hence
$\sigma_{\mathrm{adv}}(\mu,r,r^\prime,z) = -\sigma(-\mu,r,r^\prime,z)$. 
By periodicity of $\Kret$ we thus infer the important relation
\begin{equation}\label{KadvKret}
\Kadv(\mu,r,r^\prime,z) = \Kret(2\pi -\mu,r,r^\prime,z)\:.
\end{equation}

\nextparagraph
\paragraph{Symmetric solution for a point source.}

The simplest source that is compatible with helical symmetry is a
point mass $m$ in circular motion.
Let $(\barmu, \barr < \Omega^{-1}, \bar{z} =0)$ be the position of the point particle; then
the density is
\[
\rho_{\mathrm{h}}(\mu,r,z) = m \left(1 - \Omega^2 \barr^2\right)^{\frac{1}{2}} \,
\delta(\mu - \barmu) \frac{\delta(r-\barr)}{\barr} \delta(z)\:,
\]
and the associated retarded potential reads
\[
\Phiret(\mu,r,z) = m \left(1 - \Omega^2 \barr^2\right)^{\frac{1}{2}} \Kret(\mu-\barmu,r,\barr,z)\:.
\]
The radial component of the force at $(\mu,r,z) \neq (\barmu,\barr,\bar{z})$
is given by $\partial_r \Phiret$;
when $r=\barr$, $z=0$
it simplifies to
\[
\big[\partial_r \Phiret\big]\big|_{r=\barr} =
\frac{m}{2} \left(1 - \Omega^2 \barr^2\right)^{\frac{1}{2}}
\,\partial_{\barr}\, \Kret(\mu-\barmu,\barr,\barr,0)\:,
\]
where we have used the symmetry of the retarded kernel in $r$ and $\barr$, i.e.,
$\Kret(\mu,r,\barr,z) = \Kret(\mu,\barr,r,z)$.
Consequently, the fields $\Phiret$ and $[\partial_r \Phiret]|_{r=\barr}$
at positions $\mu\neq \barmu$, $r=\barr$, $z=0$
are completely described by $\Kret(\mu,\barr,\barr,0)$ and its derivatives.

To obtain the kernel $\Kret(\mu,\barr,\barr,0)$ 
we first investigate the retarded angle $\sigma(\mu, \barr, \barr,0)$.
It ensues from~\eqref{retangle} that
$\mu = 0$ corresponds to $\sigma =0$ and thus $\mu =2 \pi$ to $\sigma = 2 \pi$
by~\eqref{sigmaperiod}.
Hence, for $\mu\in[0,2\pi)$, $\sin \frac{\sigma}{2}$ is non-negative and
the defining equation for $\sigma(\mu,\barr,\barr,0)$ thus  becomes
\begin{equation}\label{newretangle}
\mu - \sigma + 2 \Omega \barr \sin \frac{\sigma}{2} = 0\:.
\end{equation}
We find $\sigma = \mu$ in the limit $\barr \rightarrow 0$;
for $\barr > 0$, however, $\sigma > \mu$, unless $\sigma = 0 =\mu$
or $\sigma = 2\pi = \mu$; the difference between the angle $\mu$ and
the retarded angle $\sigma$ is largest when $\sigma = \pi$.
Keeping $\mu \in (0,2 \pi)$ fixed (so that $\sigma$ is in the same interval)
we obtain
\begin{equation}\label{newretangledr}
\partial_{\barr} \sigma = \frac{2 \Omega \sin\frac{\sigma}{2}}{1 - \Omega \barr \cos \frac{\sigma}{2}} > 0 \:;
\end{equation}
hence, in addition to being increasing in $\mu$,
$\sigma(\mu,\barr,\barr,0)$ is increasing, with range $(\mu, \sigma_1(\mu) < 2\pi)$,
also when viewed as a function of $\barr$.

Finally, the results~\eqref{newretangle} and~\eqref{newretangledr} lead to
\begin{align}
\label{Kretbarr}
& \Kret(\mu,\barr,\barr,0) = - \frac{1}{2 \barr \sin\frac{\sigma}{2}} \;\frac{1}{1 - \Omega \barr \cos\frac{\sigma}{2}} \\[1ex]
\nonumber
& \partial_{\barr}\, \Kret =
\frac{(1- \Omega \barr)^2 + 4 \Omega \barr \sin^2\frac{\sigma}{4}}{2 \barr^2 \sin\frac{\sigma}{2}}
\Big(\frac{1}{1 - \Omega \barr \cos\frac{\sigma}{2}}\Big)^3.
\end{align}
The kernel
$\Kret$ is manifestly negative and monotonically increasing in $\barr$, since
$\partial_{\barr}\Kret$ is positive; for $\barr\rightarrow 0$ both expressions
diverge, while the limit is finite for $\barr\rightarrow \Omega^{-1}$.

Since $\Kret(\pi,\barr,\barr,0) = \Kadv(\pi,\barr,\barr,0)$ by~\eqref{KadvKret},
the retarded and advanced potential at the antipodal point $(\mu = \barmu + \pi, r= \barr, z = 0)$
are equal; the same is true for the radial components of the forces.
The tangential components, 
however, are not equal, but opposite,
since $\partial_\mu \Kret|_{\mu = \pi} = -\partial_\mu \Kadv|_{\mu=\pi}$ by~\eqref{KadvKret}.
Therefore, if we consider the \textit{symmetric solution} $\Phi(\mu,r,z)$
of~\eqref{waveeq}, i.e.,
\begin{equation}\label{mixedpot}
\Phi(\mu,r,z) = \frac{1}{2} \:\Big[\Phiret(\mu,r,z) + \Phiadv(\mu,r,z) \Big] \:,
\end{equation}
then the $\mu$-derivatives of the two terms cancel, so
that $\partial_\mu \Phi = 0$ at the antipodal point of a point mass.
This fact is a necessary prerequisite for a system of two (or more) particles
in circular motion to be in equilibrium.
Henceforth we only consider symmetric potentials~\eqref{mixedpot}.

\nextparagraph
\paragraph{Equilibrium configuration for $N$ point masses.}

%
\begin{figure}[tbp]
\begin{center}
\psfrag{r}[lc][cc][1.2][0]{$\barr_{\mathrm{e}}$}
\psfrag{m}[lc][cc][1.2][0]{$\barmu_0 =0$}
\psfrag{b}[lc][cc][1][0]{$\barmu_1 = 2 \pi/N$}
\psfrag{c}[lc][cc][1][0]{$\barmu_{N-1}$}
\includegraphics[width=0.28\textwidth]{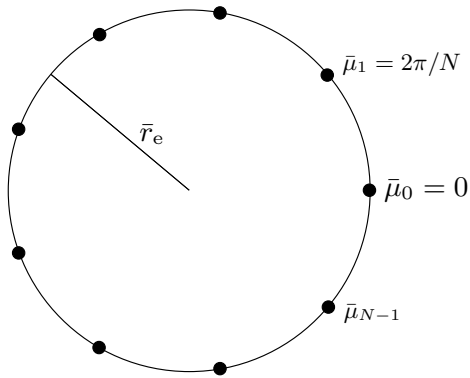}
\caption{$N$ point particles of equal mass $m$ in uniform circular motion ---
the mutual interaction is given by
the symmetric potential (half-retarded $+$ half-advanced potential).
We prove that there exists a unique radius $\barr_{\mathrm{e}}$ such that
the configuration is in equilibrium.}
\label{fig}
\end{center}
\end{figure}

We now consider the helical configuration depicted in Fig.~\ref{fig}:
let $n=0,1,\ldots,(N-1)$ be $N$ point masses of equal mass $m$,
equidistantly distributed along a circle of radius $\barr$ at $z =0$ and
uniformly rotating about their common center ---
the $n$\raisebox{0.6ex}{{\footnotesize th}} particle's
position is thus given by $(\barmu_n, \barr,0)$ with $\barmu_n = 2\pi n/N$.
Let $\Phi_n(\mu,r,z)$ denote the symmetric potential generated by
the $n$\raisebox{0.6ex}{{\footnotesize th}} particle.
At the position $(\mu,r,z) = (0,\barr,0)$ of the first point mass the
total potential $\Phi$ is then given as $\sum_{n\geq1} \Phi_n(0,\barr,0)$, hence
\[
\Phi(0,\barr,0) =
\frac{m}{2} \left(1 - \Omega^2 \barr^2\right)^{\frac{1}{2}}
\sum_{n=1}^{N-1} (\Kret+\Kadv)(\barmu_n,\barr,\barr,0).
\]
Making use of~\eqref{KadvKret} results in
\begin{align}
\nonumber
& \Phi(0,\barr,0) = m \left(1 - \Omega^2 \barr^2\right)^{\frac{1}{2}} \sum_{n=1}^{N-1} \Kret(\barmu_n,\barr,\barr,0) \\
\label{radforceN}
& \big[\partial_r  \Phi\big]\big|_{\barr} =
\frac{m}{2} \left(1 - \Omega^2 \barr^2\right)^{\frac{1}{2}}
\sum_{n=1}^{N-1} \partial_{\barr} \Kret(\barmu_n,\barr,\barr,0)
\end{align}
for the potential and the radial component of the force at $(0,\barr,0)$.
The tangential component of the force, i.e., $\partial_\mu \Phi(\mu,\barr,0)|_{\mu=0}$,
vanishes, since
$\partial_\mu \Kadv|_{\mu} = - \partial_\mu \Kret |_{2\pi-\mu}$ and
thus $\partial_\mu \Kadv|_{\mu_n}= -\partial_\mu \Kret|_{\mu_{N-n}}$; likewise,
$\partial_z \Phi(0,\barr,z) |_{z=0} = 0$, which is a simple consequence of the
mirror symmetry in $z$.
The equation of motion~\eqref{equicond} for the first particle thus reduces
to
\begin{equation}\label{equicondagain}
\big[\partial_r \Phi(0,r,0)\big]\big|_{r=\barr} - \big(1
+\Phi(0,\barr,0)\big) \: \frac{\Omega^2 \barr}{1-\Omega^2 \barr^2} =
0 \:,
\end{equation}
where we have used the independence of (proper) time of $\Phi$ at the particle's position
and $m \ddot{\bar{x}}_0^{\,\alpha} = (\partial^\alpha \xi^2)/(2 \xi^2)$, where $\xi^2 = -1 +\Omega^2 r^2$,
for the centrifugal term.
The equations for the remaining $(N-1)$ particles are identical, since the
symmetry of the configuration entails that none of the particles is distinguished.
Hereby, the system of $3 N$ equations~\eqref{equicond} reduces to one single
equation~\eqref{equicondagain}.

We conclude that the configuration of Fig.~\ref{fig} is in equilibrium,
if condition~\eqref{equicondagain} holds, i.e.,
if the radial force acting on each
particle is balanced by the centrifugal force.
As follows from~\eqref{Kretbarr}, the potential $\Phi(0,\barr,0)$
is negative and monotonically increasing for all $\Omega \barr <1$;
it diverges for $\barr\rightarrow 0$ and converges to zero when $\Omega \barr \rightarrow 1$.
Consequently, there exists a unique radius $\barr_0$ such that
$(1+\Phi(0,\barr,0))$ is negative for all $\barr<\barr_0$,
and positive for $\barr>\barr_0$.
The term $\big[\partial_r \Phi(0,r,0)\big]|_{r=\barr}$ 
is positive for all $\Omega \barr <1$; it diverges as $\barr\rightarrow 0$
and it goes to zero when $\Omega \barr \rightarrow 1$.
Combining the results it follows that the function on the
l.h.s.\ of~\eqref{equicondagain}
is positive for all $\barr\leq \barr_0$
and goes to $-\infty$ as $\Omega \barr \rightarrow 1$.
We thus conclude that this function assumes the value zero
at least once in the interval $\barr \in (\barr_0, \Omega^{-1})$,
so that there exists at least one radius $\barr_{\mathrm{e}}$ for which
condition~\eqref{equicondagain} is satisfied and the configuration is
in equilibrium.
In the following we prove that radius $\barr_{\mathrm{e}}$ is unique
by showing that the l.h.s.\ of~\eqref{equicondagain}
is decreasing for $\barr\in(\barr_0,\Omega^{-1})$.


The proof would be trivial if the radial force $\big[\partial_r \Phi(0,r,0)\big]|_{r=\barr}$
were decreasing in $\barr$
(since the second term on the l.h.s.\ of~\eqref{equicondagain} is
manifestly decreasing for $\barr > \barr_0$). However,
whether monotonicity of $\big[\partial_r \Phi(0,r,0)\big]|_{r=\barr}$
actually holds, is unclear in general.
Namely, it can be shown numerically that
$\psi(\mu,\barr) = \partial_{\barr}[(1 - \Omega^2 \barr^2)^{1/2} \partial_{\barr} \Kret(\mu,\barr,\barr,0)]$
does not have a sign: there exists a connected domain $D$ in
the set $(0,2\pi) \times (0,\Omega^{-1})$ such that
$\psi$ is positive when $(\mu,\barr) \in D$ and
negative when $(\mu,\barr) \not\in \bar{D}$ ---
this is in stark contrast to the Newtonian case, where
$\psi$ is negative for all $(\mu,\barr)$. 
The main properties of $D$ are the following:
$\min_D \barr \approx 3/4 \:\Omega^{-1}$, hence
negativity
holds for small $\barr$, where velocities are small compared to $c$
so that Newtonian gravity is a good approximation
to scalar gravity;
$\max_D \mu \approx 1/4$, hence
the radial force is decreasing at least when the number of
particles is sufficiently small, i.e.,
when $\barmu_n > 1/4$ $\forall n\,$;
typical values of $\psi$ on $D$ are by several orders of magnitude larger than
typical values of $|\psi|$
on $\big[(0,2\pi) \times (0,\Omega^{-1})\big] \backslash \bar{D}$ ---
this complicates matters when one seeks to prove
that the sum over all $\barmu_n$ is negative.
Despite this last remark,
numerical evidence suggests that $\big[\partial_r \Phi(0,r,0)\big]|_{r=\barr}$
is in fact decreasing irrespective of the number of particles;
a rigorous proof, however, seems difficult to obtain.

In our proof we therefore proceed along different lines.
The derivative of the function on the l.h.s.\ of~\eqref{equicondagain}
reads
\begin{align}
\nonumber
m & \sum_{n=1}^{N-1} \Bigg\{
\partial_{\barr} \Big[\frac{1}{2}  \left(1 - \Omega^2 \barr^2\right)^{\frac{1}{2}}
\partial_{\barr}\, \Kret(\barmu_n,\barr,\barr,0) \Big]  \\
\nonumber & \qquad - \, \partial_{\barr} \left[ \left(1 - \Omega^2
\barr^2\right)^{\frac{1}{2}} \Kret(\barmu_n,\barr,\barr,0) \right]
\frac{\Omega^2 \barr}{1-\Omega^2 \barr^2}
\Bigg\}\,  \\
\label{entireder}
&  \: - \: \big(1 +\Phi(0,\barr,0)\big)\:
\partial_{\barr} 
\left(\frac{\Omega^2 \barr}{1-\Omega^2 \barr^2}\right)\:.
\end{align}
Since
the last line is clearly negative when $\barr > \barr_0$,
in order to show that the whole function is negative, it suffices to prove that
each of the terms in braces is negative individually.
To this end let
$\sigma = \sigma(\barmu_n,\barr,\barr,0)$ for some $n$;
then each individual term in braces has the form
\begin{equation}\label{eachsummand}
\frac{\left(1 - \Omega^2 \barr^2\right)^{- \frac{1}{2}}}%
{4 \barr^3 \left( 1- \Omega^2 \barr^2 \cos \frac{\sigma}{2} \right)^5 \sin \frac{\sigma}{2}}\:\,
P(\Omega \barr, \cos\frac{\sigma}{2}) \:,
\end{equation}
where $P(\barv,\cos \frac{\sigma}{2})$ is a complicated polynomial of degree eight in $\barv = \Omega \barr <1$
and of degree four in $\cos\frac{\sigma}{2}$.
In a second step we replace $\cos \frac{\sigma}{2}$ by a variable $\delta$ defined through
$\cos \frac{\sigma}{2} = \barv^{-1} [1 - (1-\barv^2) \delta]$.
Since $-1 \leq \cos\frac{\sigma}{2} \leq 1$, the permitted range of $\delta$ is
\begin{equation}\label{deltarange}
\frac{1}{2} < \frac{1}{1+\barv} \leq \delta \leq \frac{1}{1-\barv}\:.
\end{equation}
Using $\delta$ leads to
a simple representation of $P(\barv, \cos\frac{\sigma}{2})$:
\[
P = -\frac{ 3 - 8 \delta  +  \delta^2 (5 - 4 \barv^2)
+ \delta^3 (2 + 4 \barv^2) + 2 \delta^4 \barv^4}{(1-\barv^2)^{-4}}\:.
\]
The roots of the polynomial $P$ are explicitly given by
$\barv^2 = \delta^{-2} ( 1 - \delta \pm \sqrt{\Delta})$, where
the discriminant $\Delta$ reads
\begin{equation}
\Delta =
-2 \left(\delta-\frac{1}{2}\right) \left(\delta - [\sqrt{2}-1] \right)\left(\delta + [\sqrt{2}+1]\right)\:.
\end{equation}
Evidently, $\Delta$ is non-negative if and only if $\delta \leq -1-\sqrt{2}$ or
$\delta \in [\sqrt{2}-1,\frac{1}{2}]$. As a consequence, the roots of $P$
lie outside of the admissible domain~\eqref{deltarange} of the variables $(\barv,\delta)$.
Since in addition $P<0$ for $\barv \rightarrow 0$ and $\delta =1$, it follows
that $P$ is negative everywhere on the admissible $(\barv,\delta)$-domain, or, equivalently,
\begin{equation}
P(\Omega \barr, \cos\frac{\sigma}{2}) < 0 \qquad \forall (\barr,\sigma) \in (0, \Omega^{-1}) \times [0,2 \pi) \:.
\end{equation}
With $P<0$ the expression~\eqref{eachsummand} is negative,
which completes the proof of the claim.


\nextparagraph
\paragraph{Post-Newtonian expansion.}

For a given number of particles,
the equilibrium radius $\barr_{\mathrm{e}}$ of the con\-figuration in Fig.~\ref{fig}
is a function of the angular velocity $\Omega$ and the mass.
This functional dependence cannot be made explicit,
since this would involve, among other things, an explicit knowledge of
the retarded angle~\eqref{newretangle}.
(For a two-particle system, $\barr_{\mathrm{e}}$ can be given as
a (non-explicit) function of the orbital
velocity $\Omega \barr_{\mathrm{e}}$, 
see~\cite{Bruneton:2006}, which, of course,
does not lead to an explicit solution for $\barr_{\mathrm{e}}$.)

It is feasible, however, to analyze the equilibrium condition~\eqref{equicondagain}
by means of a post-Newtonian approximation scheme.
With the support of a computer algebra program necessary manipulations
can be done in a straightforward way
and we eventually obtain a post-Newtonian expansion of $\barr_{\mathrm{e}}$;
here, we merely state some results.

Let $\omega$ be the angular velocity as measured in standard units,
i.e., $[\omega] = s^{-1}$; clearly, $\omega = \Omega c$, where $c$
is the speed of light; furthermore, let $G$ be the gravitational constant
and $M = N m$ the total mass of the $N$-particle configuration of~Fig.~\ref{fig}.
We define a quanitity $R$ (with unit length) and a dimensionless
quantity $x$ according to
\[
R =  \big(G M \omega^{-2}\big)^{\frac{1}{3}}\:,
\qquad
x = \big( G M \omega c^{-3} \big)^{\frac{1}{3}}\:.
\]
In terms of $R$, $x$ the post-Newtonian expansion of
$\barr_{\mathrm{e}}$ is:

\vspace{-4ex}

\begin{center}
\begin{tabular}[t]{|c|c|}
\hline $N$ & $\barr_{\mathrm{e}}$ \\ \hline $2$ & $\frac{1}{2} R\:
\left[ 1 + x^2/12 - 7 x^4/72 + O(x^6) \right]$ \\
$3$ & $\frac{1}{\sqrt{3}} R \:
\left[ 1 + 7 x^2/72 - 0.065 x^4 + O(x^6)\right]$ \\
$\vdots$ & $\vdots$ \\
$10^5$ & $1.228 R \:\left[ 1 + 0.273 x^2 - 0.465 x^4 + O(x^6) \right]$ \\
\hline
\end{tabular}
\end{center}

For highly relativistic configurations numerical investigations
indicate that $\barr_{\mathrm{e}} \propto \Omega^{-1}$ as $\Omega
\rightarrow \infty$, so there does not exist an innermost circular
orbit.


\newpage

\begin{acknowledgments}
We would like to thank J.\ Bi\v{c}ak for useful discussions
and the Isaac Newton Institute 
(Cambridge) for 
support.
\end{acknowledgments}




\end{document}